# Free energy profile computations for force-induced melting of a short B-DNA using Jarzynski equality

S. Siva Nasarayya Chari[1,*] and Prabal K. Maiti[2,†]

[1]*Department of Physics, Faculty of Science, University of Allahabad, Prayagraj-211002, Uttar Pradesh, India.*

[2]*Department of Physics, Centre for Condensed Matter Theory, Indian Institute of Science, C V Raman Ave, Bengaluru-560012, Karnataka, India.*

**Abstract**

We initially equilibrate a system of solvated 12-base-pair canonical B-DNA of a specified sequence in a state of no external force acting on it (state A: $f = 0$). A switching experiment is then performed by pulling one end of the DNA while restraining its other end. The pulling force is taken up to a maximum of 400 pN (state B), at a specified constant rate. The non-equilibrium work done $W$ associated with this finite-time process is determined by integrating the extension of the DNA molecule with respect to the applied force. The distribution of the work values $P(W)$ is obtained by repeating the pulling process several times, wherein each time, the process is started with macroscopic state A. The free energy profile $\Phi(f)$ is obtained by a suitably devised method from the Jarzynski equality. The conformational space of the DNA is also sampled in an equilibrium method of umbrella sampling by considering the end-to-end length $\xi$ as the reaction coordinate. The associated potential of mean force (PMF) $\Phi(\xi)$ is then computed by invoking the weighted histogram analysis method (WHAM). From the results, we observe that the equilibrium free energy difference between two states, A and B, $\Delta F_{AB} (= F_B - F_A)$ obtained through both methods are in close agreement with each other, despite the differences along the profiles $\Phi(f)$ and $\Phi(\xi)$, attributed to their distinct pathways of unzipping.

*Keywords*: Force-induced melting of DNA, Jarzynski equality, single-molecule pulling experiments, umbrella sampling.

## 1 Introduction

Fundamental processes that occur at the level of nucleosomes are connected to the denaturation and replication of the double-stranded DNA (dsDNA) by proteins like helicase, histone chaperones, or transient thermal fluctuations. The associated free energy profiles along the pathways of denaturation at various physiological conditions of the dsDNA are of utmost importance to the pharmaceutical and drug manufacturing sectors. A dsDNA can either be heated or be subjected to an external mechanical force for the denaturation to occur, the former method is known as the thermal melting [1–3], whereas the latter is termed as the force-induced melting [4–6] of the DNA. In vitro, such a process is successfully performed through force spectroscopy by AFM cantilever [7–9] or by the applications of optical tweezers [10–12]. In silico, the model systems [13–17], and atomistic simulations have been playing a vital role in free energy calculations pertaining to the biomolecular systems like DNA, and RNA [11,18–22], and their interactions with various proteins [23–26].

---

* snchari@allduniv.ac.in (corresponding author)

† maiti@iisc.ac.in

Equilibrium simulation methods like umbrella sampling [27,28], and thermodynamic integration [29] have been the conventional approaches for free energy and PMF profile calculations for biological systems ranging from ion channels in membranes [30,31], protein-ligand interactions [32,33], and water permeation through Carbon nanotube [34,35] are a few to name. These methods were considered a benchmark in their regime. However, in the early twenty first century, it has seen an increasing number of investigations over the nonequilibrium fluctuation theorems, particularly the Jarzynski equality [36,37], Crooks' fluctuation theorem [38], and their applicability to biological systems for their free energy computations [19,39,40]. Soon these nonequilibrium methods have become increasingly popular and are validated for various physical [41–43], chemical and simple biological systems [39,43–47]. However, their applicability for complex biomolecular systems had been a debated topic in the literature.

Warshel et. al. [48] conducted the simulations of proteins and observed the difference in the electrostatic potential values obtained from nonequilibrium techniques like JE, and the free energy perturbation techniques like US. In another study, Kuyucak et. al. [49] performed the PMF calculations of a ligand binding to an ion channel, obtained from both JE and US; wherein, they reported a slow convergence in the PMF obtained from JE. However, Shulten et. al. [19,50] conducted the steered molecular dynamics simulations over a deca-alanine system. From their PMF calculations from JE, and US, almost equal efficiency is reported between both the methods, and an improving accuracy with the sampling size in JE. Also, other work by Ritort et. al. [39,46] report the viable application of the nonequilibrium fluctuation theorems to the biomolecular systems like DNA hairpin.

Therefore, keeping in view of the significance of the free energy calculations corresponding to the DNA, here we compare the respective PMF calculations from both the equilibrium and nonequilibrium methods. Pertaining to the calculations for a dsDNA, Maiti et. al. [51,52] have reported the nonequilibrium pulling processes carried out with various possible pulling protocols. However, a conventional umbrella sampling had been employed for free energy calculations whenever required, to determine the corresponding PMF profile. Here, in this work, we perform nonequilibrium switching experiments over an initially equilibrated dsDNA and employ the Jarzynski equality to evaluate the free energy profile along the path of the control variable, namely the pulling force. We also perform the equilibrium umbrella sampling over the same system of an equilibrated dsDNA and then determine the PMF profile. The profile from JE is then validated with the one obtained from US, after suitable normalization of their range of values. From the results we notice that the value of the free energy difference $\Delta F$ between the initial equilibrated state ($A$) and the final denaturized state ($B$) agrees well between both the methods. However, a difference of more than $10k_BT$ is observed in the midrange values of the reaction coordinate, which can be attributed to the difference in the ensemble variables, in both the cases. More details on the theory and the methods of implementation are discussed in the following sections.

## 2 Theory and Methods

The initial structure of a 12-base-pair double-stranded DNA with the sequence *CGCGAATTCGCG* is prepared using the LEaP module of AMBER-10 [53] with *parmbsc0* force field parameters describing the interactions associated with the DNA. The $Na^+$ counterions with Joung-Cheatham ion parameters [53] are added to neutralize the negative charge on the DNA. The resulting structure is then solvated with TIP3P [54] water molecules in a large enough box, ensuring a solvation shell of about 10Å even in the fully stretched configuration of the DNA. The system is then well equilibrated at a temperature of 300 K, as per the standard equilibration protocol [51,52]. The equilibrated dsDNA is then subjected to a pulling force on the O3' and O5' atoms of one of its ends, where the other end is fixed by a strong positional restraint.

### A. Jarzynski Equality

Consider a system in thermal equilibrium with a heat reservoir at a temperature $T$. Let $\lambda$ be any externally controllable macroscopic parameter tuning for which the state of the system can be changed. If the system is driven isothermally from an initial equilibrium state $A$ (consistent with $\lambda = \lambda_i$) to state $B$ (consistent with $\lambda = \lambda_f$) *via* a predefined protocol $\lambda(t)$: $\lambda_i \rightarrow \lambda_f$; then, the work done during such a switching process is

$$W = \int_0^{\tau} \frac{\partial H}{\partial \lambda} \frac{\partial \lambda}{\partial t} dt. \tag{1}$$

Here, $H$ defines how the energy of the system depends on the parameter $\lambda$, and $\tau$ refers to the switching time, indicating how slow or fast the switching process is carried out. One obtains an ensemble of work values $\{W_i\}$ by performing several such switching experiments following the same protocol $\lambda(t)$. Jarzynski equality [36,37] relates the non-equilibrium work done and the equilibrium free energy difference $\Delta F (= F_B - F_A)$ between the two states $A$ and $B$.

$$\left\langle e^{-\beta W} \right\rangle = e^{-\beta \Delta F}, \tag{2}$$

where $\langle \cdot \rangle$ denotes averaging over the work ensemble, and $\beta = (k_B T)^{-1}$. In this work, we consider the pulling force $f(t)$ as the control parameter.

### B. Umbrella sampling

In this method, one usually defines a generic variable named the reaction coordinate $\xi$ which can be a single parameter or a combination of two or more; the latter is termed as a 'collective variable.' For a process like stretching a double-stranded DNA, the end-to-end distance (EED) is chosen as the suitable reaction coordinate. In order to sample the configuration space of the DNA at any given EED $\xi_0$ adequately well, a harmonic biasing potential $U_b(\xi(r)) = k(\xi(r) - \xi_0)^2$ is applied to the system [27,28], where $\xi$ and $k$ denote the instantaneous value of EED and the steepness of the potential respectively.

Let $E_u = E(r)$ and $E_b = E(r) + U_b(\xi(r))$ representing the unbiased and biased energies of the system respectively. The equilibrium probability distribution of EED for the unbiased and biased cases separately can be written as

$$P_u(\xi) = \frac{\int e^{-\beta E_u} \delta(\xi(r) - \xi) d^N r}{\int e^{-\beta E_u} d^N r}, \tag{3}$$

and

$$P_b(\xi) = \frac{\int e^{-\beta[E_u + U_b(\xi(r))]} \delta(\xi(r) - \xi) d^N r}{\int e^{-\beta[E_u + U_b(\xi(r))]} d^N r}, \tag{4}$$

respectively, with $N$ representing the number of degrees of freedom for the system. By integrating the above with reference to the Dirac-delta function, one can obtain

$$P_b(\xi) = P_u(\xi) e^{-\beta U_b(\xi)} \frac{\int e^{-\beta E(r)} d^N r}{\int e^{-\beta[E(r) + U_b(\xi(r))]} d^N r} \tag{5}$$

which can be rewritten as

$$P_u(\xi) = P_b(\xi) e^{\beta U_b(\xi)} \left\langle e^{-\beta U_b(\xi(r))} \right\rangle. \tag{6}$$

The unbiased free energy associated with $\xi(r) = \xi$ can be obtained from $P_u(\xi)$ as,

$$\begin{aligned} A_u(\xi) &= -k_B T \ln(P_u(\xi)), \\ &= -k_B T \ln(P_b(\xi) - U_b(\xi)) + F, \end{aligned} \tag{7}$$

where $F = -k_B T \ln\langle e^{-\beta U_b(\xi(r))} \rangle$. The reaction coordinate $\xi$ is sampled at close enough regular discrete points so that $P_b(\xi)$ for neighboring values of $\xi$ has significant overlap. But to obtain the potential of mean force (PMF) along the path of $\xi$, we need to stitch the values of $A_u^i(\xi)$ that were obtained at each discrete point of $\xi$, which requires an accurate evaluation of $F$ at each window. Though a few methods [55,56] were proposed for this purpose, the most promising one is the weighted histogram analysis method (WHAM) [57].

## 3 Results and Discussion

An initial configuration of the ds-DNA after equilibration which corresponds to state A: $f = f_0 = 0$ pN is depicted in fig. 1(a). Now, a pulling force is applied on the O3' and O5' end atoms of the ds-DNA while applying a strong positional restraint (of force constant ~5000 N/m) on the O5' and O3' atoms of the other end of the DNA. This pulling force is applied at a constant rate of $10^{-4}$ pN/fs, until it reaches a maximum value of $f_{max} = 400$ pN. Therefore, at any instant of time $t$, the applied force is given by

$$\begin{aligned} f(t) &= f_0 + (f_{max} - f_0)\frac{t}{\tau}, \\ &= f_0 + t(\delta f_\tau). \end{aligned} \tag{8}$$

In the above, $\tau$ refers to the switching time, and $\delta f_\tau = (f_{max} - f_0)/\tau = 10^{-4}$ pN/fs which is the switching rate. Note that here the rate of applied force exceeds the typical experimental value by almost two orders of magnitude. This is because invoking the experimental pulling speeds in a simulation requires huge simulation times that are practically impossible. Figure 1(b) represents a denaturized configuration at $f = f_{max}$.

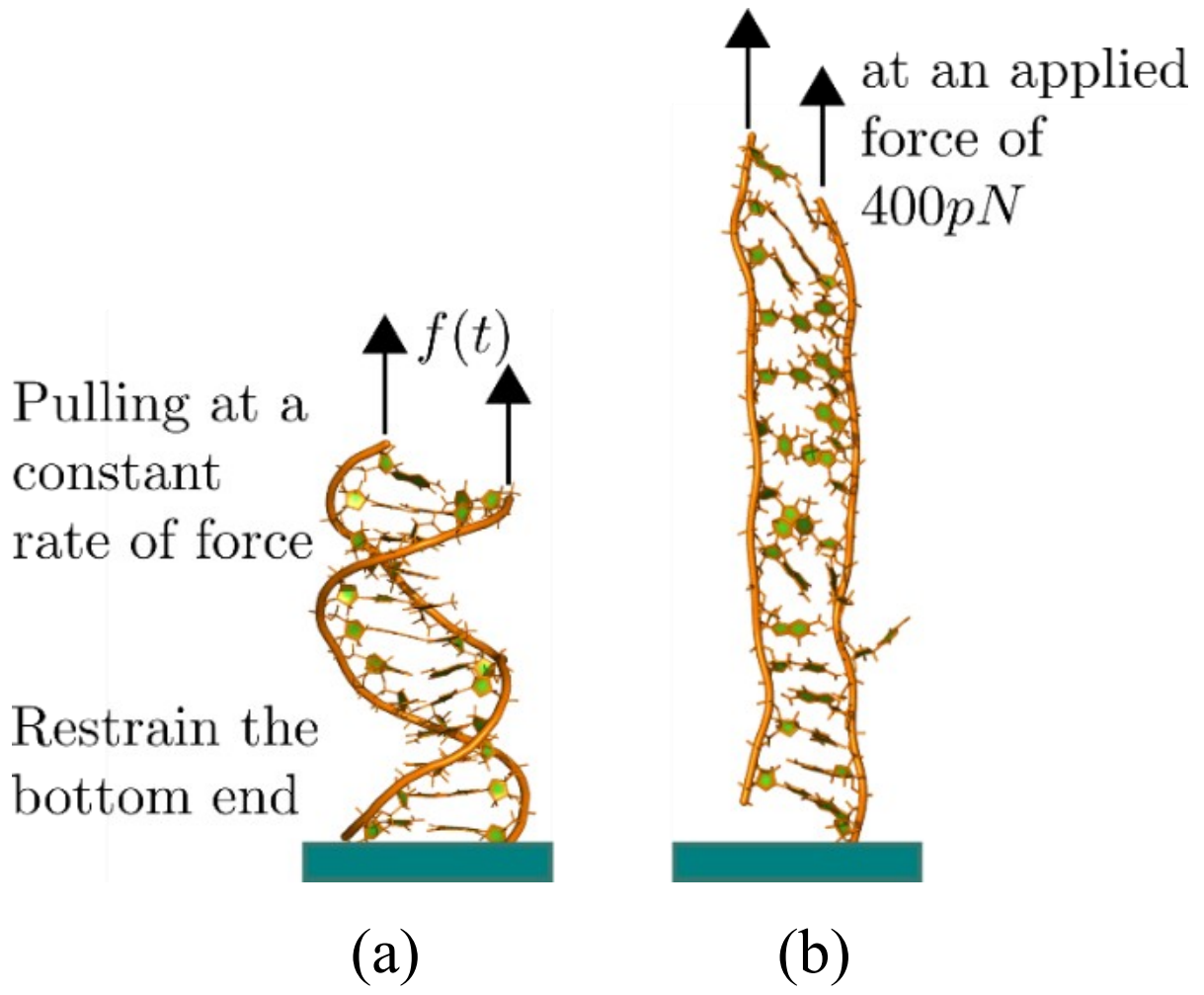

Figure 1. (a) An initial equilibrated configuration of the DNA at no pulling force. (b). a configuration of the DNA at an applied pulling force of 400 pN, over the O3' – O5' atoms of the same end.

The resulting molecular extension of the dsDNA as a function of the pulling force is shown in fig. 2(a). There exist three H-bonds between a C-G base pair and two H-bonds between a base pair A-T, respectively [58]. A H-bond is considered broken if the respective atoms are separated by more than 3.2Å of distance, and the angle is below 130°. Following this criterion, the number of Watson-Crick H-bonds that are intact during the pulling process is calculated from the trajectory and is shown in fig. 2(b).

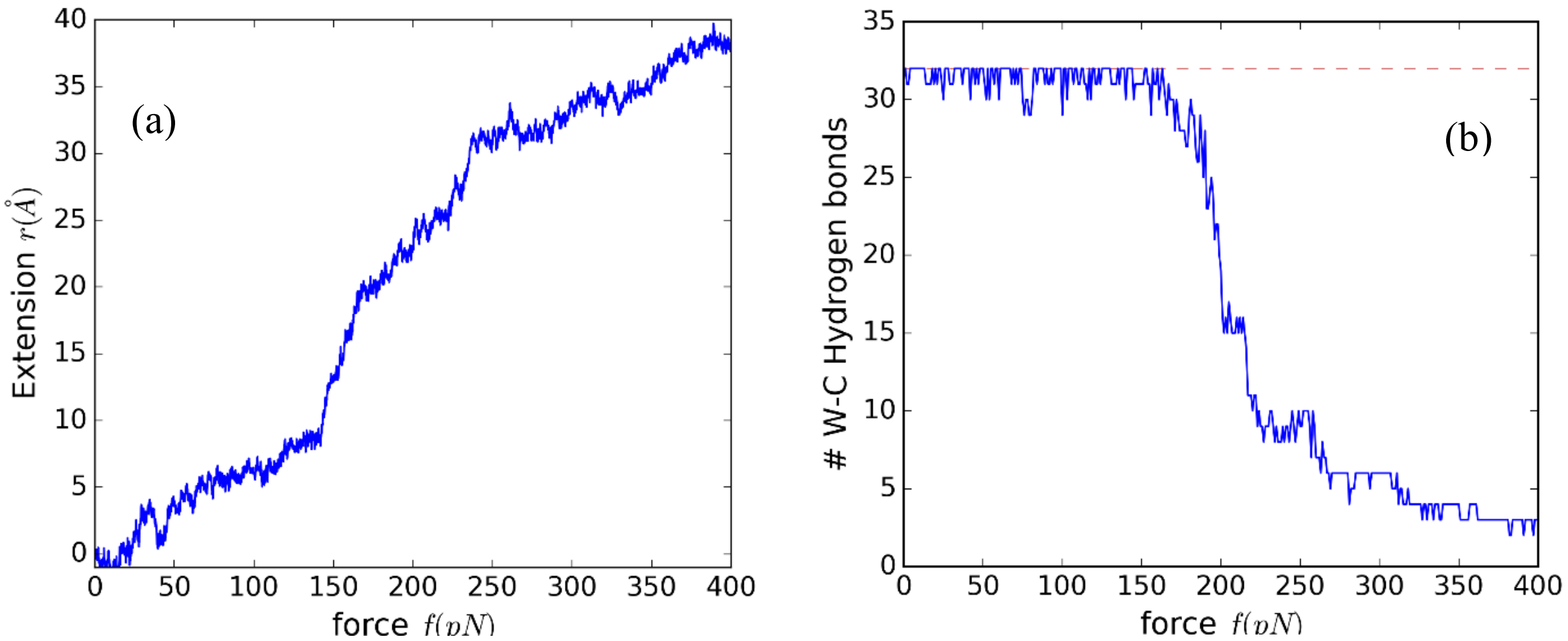

Figure 2. (a). Force-extension curve of the ds-DNA during a numerical pulling experiment (b). the number of Watson-Crick H-bonds that are intact during the process.

From the figure, we notice that most of the H-bonds were intact until about 150 pN of pulling force, beyond which, the dsDNA was tending to transit from a helical configuration to a stretched, ladder-like configuration as shown in fig. 1(b), with majority of its H-bonds broken. Since this pulling is performed as a finite time process, it results in a final state which may not be in equilibrium, but consistent with the condition that $f = f_{max}$.Therefore, the nonequilibrium work done associated with this process $A(f = 0) \rightarrow B(f = f_{max})$ can be evaluated by numerically integrating the force-extension curve,

$$W = \int_{f_0}^{f_{max}} x(f)df = \int_{0}^{400\text{pN}} x(f)df. \tag{9}$$

This corresponds to a single switching process. Several of such identical pulling experiments were performed to obtain an ensemble of work values, $P(W)$, as depicted in fig. 3.

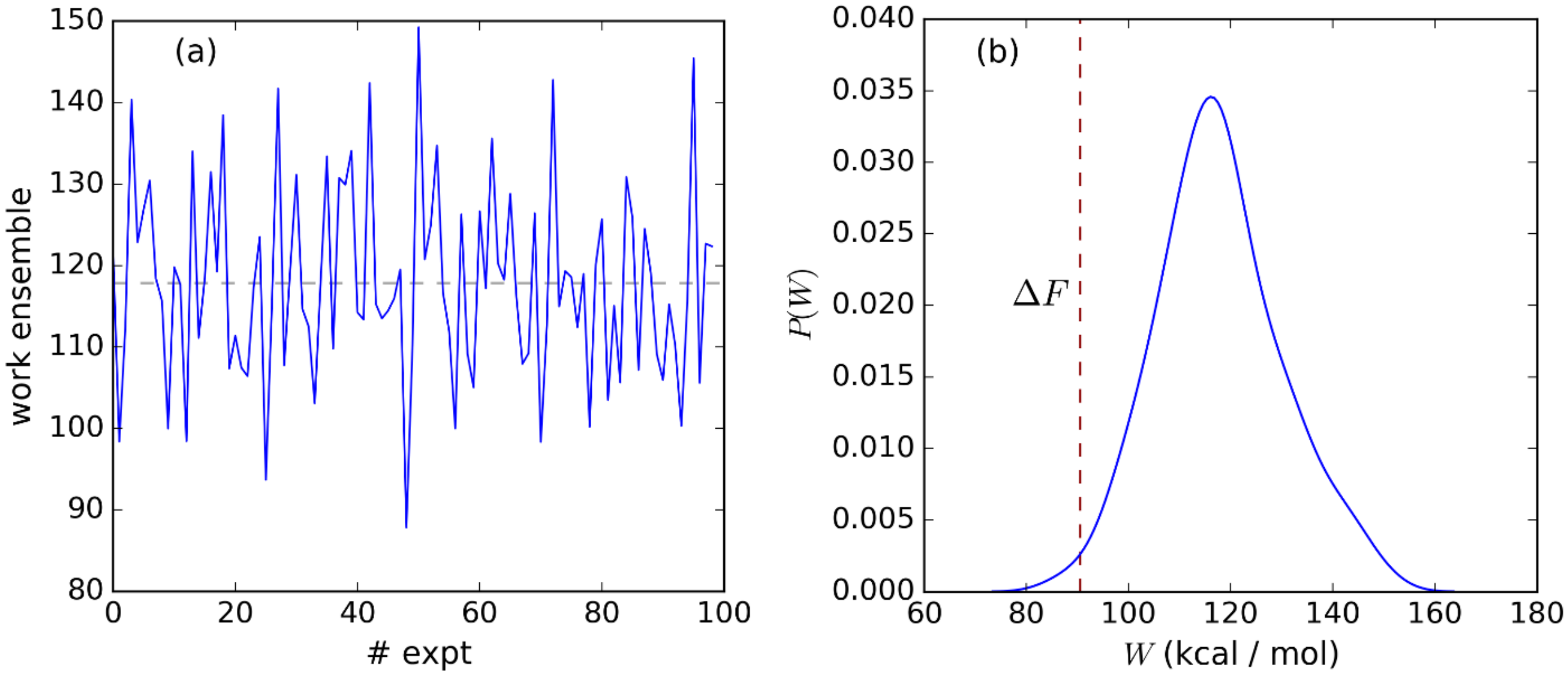

Figure 3. (a). The non-equilibrium work values $\{W_i\}$ obtained through various pulling experiments (b). normalized distribution of work done, $P(W)$. The red dotted line denotes the free energy difference $\Delta F$ calculated using the Jarzynski equality.

The free energy difference $\Delta F$ between the initial and final equilibrium states of A and B can be obtained following the Jarzynski equality, as in eq. (2). The left tail of $P(W)$, where $W < \Delta F$, represents the switching experiments that seemingly violate the second law of thermodynamics [41]. However, the mean work value $\langle W \rangle > \Delta F$, being consistent with the second law. It is straight forward to calculate the free energy difference $\Delta F$ between the two equilibrium states, using the Jarzynski equality. In this work, we obtain the free energy profile $\Phi(f)$ along the path of the control variable $f: 0 \rightarrow 400$ pN. This full range of $f$ is divided into 100 uniformly spaced intervals, each of width 4 pN. Each intermediary point in these set of values would serve as an apparent final state for the switching process. That means, for any i-th intermediary point $B_i$ between the states $A$ to $B$,

$$\begin{gathered} B_i : f_0 + i\delta f, \text{ for } i \in [0, n], \\ \text{where } f_0 + n\delta f = f_{max}, \text{ and } \Delta F_{AB_i} = F_{B_i} - F_A. \end{gathered} \tag{10}$$

In the above, $\Delta F_{AB_i}$ denote the free energy value at $B_i$ relative to its value at state $A$; which can be obtained from the Jarzynski equality,

$$\langle e^{-\beta W_{AB_i}} \rangle = e^{-\beta \Delta F_{AB_i}} ,$$
$$\Delta F_{AB_i} = F_{B_i} - F_A = -\frac{1}{\beta} \ln\left( \langle e^{-\beta W_{AB_i}} \rangle_{A \to B_i} \right). \quad (11)$$

In the above, the ensemble averaging is done only up to the intermediary point $B_i$, and $W_{AB_i}$ represent the nonequilibrium work done from $A$ to the point $B_i$.

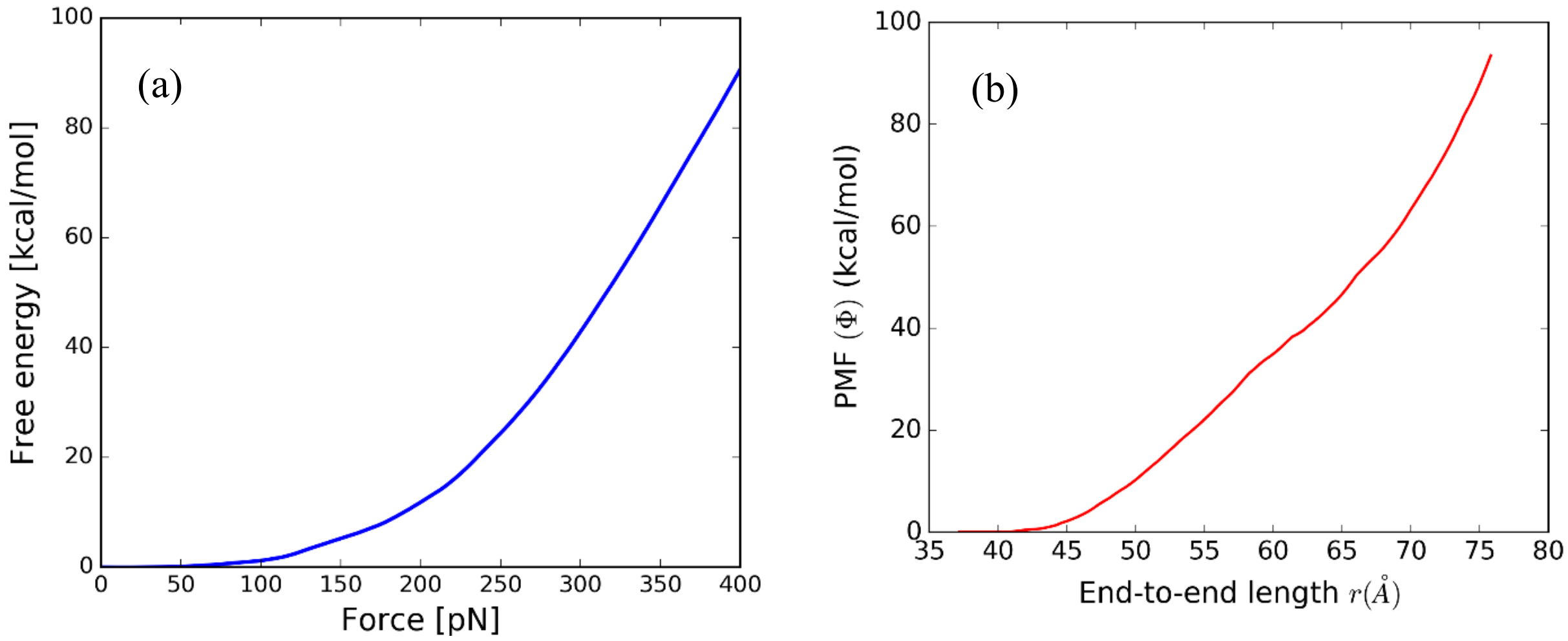

Figure 4. (a). The free energy profile $\Phi(f)$ obtained from the presented method from the Jarzynski equality. (b). The potential of the mean force (PMF) profile $\Phi(\xi)$ as obtained from the umbrella sampling method, where $\xi$ is the end-to-end length of the dsDNA.

The so obtained profile $\Phi(f)$ from this method is depicted in fig. 4(a). Point to note here is that $\Phi(f)$ at each $B_i$ represent only the free energy of $F_{B_i}$ relative to $F_A$, but not the absolute free energy.

We also performed the equilibrium simulation method that samples the configuration space of the dsDNA at several end-to-end lengths spanning from 36Å to 76Å. This is achieved by applying a harmonic potential restraint $k(\xi - \xi_0)^2$of force constant $k = 4$ kcal/mol-Å$^2$, at an initial end-to-end length value of $\xi_0 = 36$Å, which is then varied at a window of 1Å. The system is sufficiently equilibrated at each window, resulting in the Gaussian-like forms for $P(\xi)$, with sufficient overlap between the distribution tails for the neighboring windows. This combination of the force constant and the sampling window values were chosen due to their good convergence in the PMF calculations as shown in the previous studies [51,52]. The umbrella sampling method combined with the WHAM resulted in the PMF profile $\Phi(\xi)$, as shown in fig. 4(b). Here, the end-to-end length $\xi$ is chosen as the reaction coordinate. Therefore, the sequence of microstates during the denaturation of the dsDNA in this protocol can significantly differ from that of the sequence from the nonequilibrium pulling process, as the pulling force is the control variable in the latter case. However, both profiles are expected to merge in the case of reversible pulling, i.e., for large switching time limit. Despite the differences along the profile, the value of $\Delta F_{AB} = F_B - F_A =$ 91.2 kcal agrees well between both the methods.

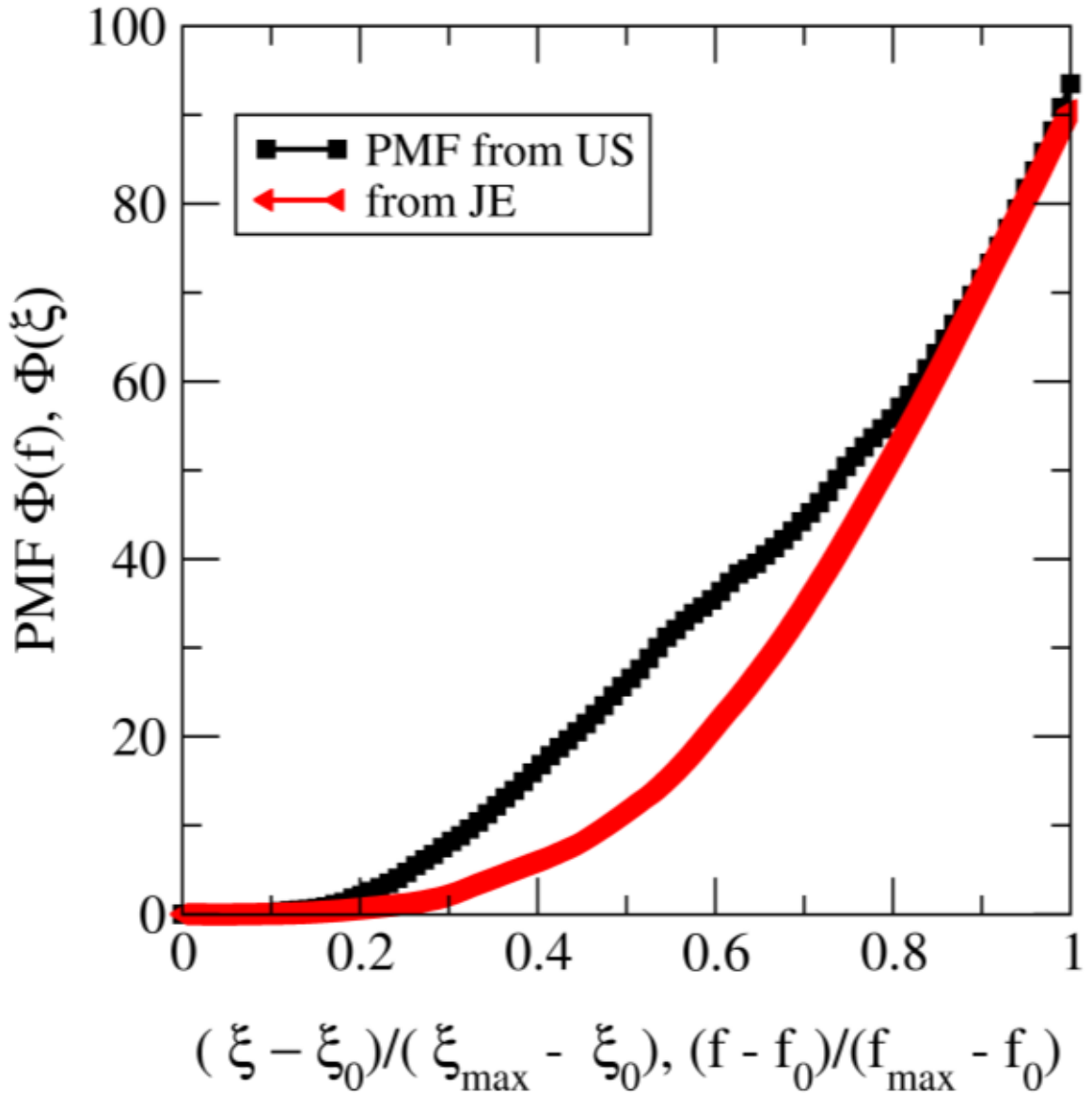

Figure 5. Profiles of free energy obtained through the Jarzynski equality, $\Phi(f)$, and through the method of umbrella sampling, $\Phi(\xi)$.

Since the range of the pulling force $f$, and the end-to-end distance $\xi$ are completely different, both are normalized to get values from 0 to 1, and the combined plot is shown in fig. 5. It is evident from fig. 5 that both methods yield almost overlapping profiles, except in the mid-range values of the normalized $\xi$ and $f$, which can be attributed to differences in their pathways of unzipping.

## 4 Conclusion

The free energy profile for the force-induced melting of a 12-base pair short B-DNA with the sequence *CGCGAATTCGCG* is obtained using the nonequilibrium work fluctuation theorem, specifically the Jarzynski equality. Being a finite time process, the final state of this pulling need not be an equilibrium state. Therefore, considering any intermediary point $B_i$ between the initial ($A$) and final ($B$) states as an apparent final state, the JE can be applied to extract the free energy difference $\Delta F_{AB_i}$ from the set of phase space trajectories from $A$ to $B$. Following this procedure, the profile $\Phi(f)$ from $f_0$ to $f_{max}$ is obtained. The same is compared with the PMF profile $\Phi(\xi)$ obtained with the equilibrium simulation method like umbrella sampling. Here, the conformation space of an identical dsDNA is sampled at various end-to-end lengths ($\xi$) from 36Å - 76Å, by applying a suitable biasing harmonic potential restraint.

The JE method uses pulling force as the macroscopic control variable to drive the state of the dsDNA, whereas the equilibrium US method uses a collective variable, namely the end-to-end length as the reaction coordinate. Therefore, they both correspond to a different ensemble altogether. However, the profile from JE is expected to tend towards the one obtained from US, in the large switching time limit. That makes the switching process very slow, and practically impossible to perform in any reasonable time. Though the JE is accurate enough even in the small switching time regimes, but it needs many sampling paths from $A$ to $B$ so that the rare events are also properly accommodated in the trajectory, which plays a crucial role in determining the accuracy of the $\Delta F$ value. Hence, it is a trade-off between the speed and accuracy. One must judiciously decide the switching rate depending on the system size. For the rate of switching ($\delta f_\tau = 10^{-4}$ pN/fs) used here, single switching experiment from $f_0$ to $f_{max}$ is roughly about 4 ns simulation time, i.e., for $N$ such identical pulling processes, the total simulation time scales as $\sim 4N$ ns. Whereas, for US, the simulation time associated with each 1Å window is typically about 3 to 4 ns, which makes the total simulation time associated with the US, to sample from 36Å to 76Å is about 120 – 160 ns. Therefore, for this system of 12-base pair dsDNA, one can employ the JE with reasonable switching rates, if the primary concern is only the free energy difference. Whereas one can adopt the US method to yield the full PMF profile. JE would be beneficial for the small system sizes. Increasing the size of the system increases the computational complexity for JE as it requires more phase space trajectories to sample from $A$ to $B$ for better accuracy. However, these simulations can be run in parallel thereby significantly reducing the overall computation time, unlike the serial runs as required by the US.

**Acknowledgments**

SSNC thanks the computational facility provided by the Department of Physics, Indian Institute of Science. The computational facility at the Solid State and Structural Chemistry Unit (SSCU), Indian Institute of Science, Bengaluru is highly acknowledged.